# Title: Direct Measurement of Exciton Dispersion in the Long-Wavelength Limit

**Authors:** Peiyi He[1,2]†, Jiade Li[1,2]†, Jiakai Wang[1,2], Jiangxu Li[3], Jiahao Wang[4], Weiyu Sun[5], Yiwen Song[6], Xiaoyue Gao[1,2], Quanlin Guo[6], Bo Han[1,2], Ruochen Shi[1,2], Niklas Dellby[7], Tracy Lovejoy[7], Xing-Qiu Chen[3], Kaihui Liu[6], Yu Ye[6], Hailin Peng[4,5], Peng Gao[1,2,8,9]*

**Affiliations:**

[1]International Center for Quantum Materials, School of Physics, Peking University; Beijing, 100871, China.

[2]Electron Microscopy Laboratory, School of Physics, Peking University; Beijing, 100871, China.

[3]Shenyang National Laboratory for Materials Science, Institute of Metal Research, Chinese Academy of Sciences; Shenyang, 110016, China.

[4]Center for Nanochemistry, Beijing Science and Engineering Center for Nanocarbons, Beijing National Laboratory for Molecular Sciences, College of Chemistry and Molecular Engineering, Peking University; Beijing, 100871, China.

[5]Academy for Advanced Interdisciplinary Studies, Peking University; Beijing, 100871, China.

[6]State Key Laboratory for Artificial Microstructure and Mesoscopic Physics and Frontiers Science Center for Nano-optoelectronics, School of Physics, Peking University; Beijing, 100871, China.

[7]Bruker AXS LLC; Kirkland, WA, 98034, USA.

[8]Interdisciplinary Institute of Light-Element Quantum Materials and Research Center for Light-Element Advanced Materials, Peking University; Beijing, 100871, China.

[9]Tsientang Institute for Advanced Study; Zhejiang, 310024, China.

† These authors contributed equally to this work.

* Correspondence address: Peng Gao (pgao@pku.edu.cn)

**Abstract**

Exciton dispersion, which governs the propagation, scattering and radiative decay of electron-hole pairs, is essential to optoelectronics and quantum materials. In two-dimensional systems, weakened dielectric screening and long-range electron-hole exchange are predicted to induce nonanalytic exciton dispersion in the long-wavelength limit. However, direct quantitative characterization of its dimensional evolution remains lacking, especially in the ultralow-$q$ regime ($q$ < 0.02 Å$^{-1}$). Here we employ defocus-engineered momentum-resolved electron energy-loss spectroscopy in scanning transmission electron microscopy, achieving an ultrahigh momentum resolution of 0.0002 Å$^{-1}$. Using freestanding hBN as a prototypical platform, we resolve layer-dependent exciton dispersion and quantify its characteristic crossover momentum and group velocity in the long-wavelength limit. With increasing thickness, the nonanalytic linear-dispersion regime is progressively compressed, manifested by a reduction in characteristic crossover momentum $q_c$ from $1.82 \times 10^{-1}$ Å$^{-1}$ in the monolayer to $3.0 \times 10^{-3}$ Å$^{-1}$ in 25 layers. Meanwhile, the low-$q$ group velocity increases from $2.0 \times 10^{-3}\ c$ to $2.9 \times 10^{-2}\ c$, before the dispersion ultimately approaches the bulk-like parabolic limit. We further examine how the exciton band structure of monolayer hBN responds to its surrounding environment, including temperature, adjacent graphene layers, and interlayer twist in BN/graphene heterostructures. These findings uncover the fundamental physics of low-dimensional excitons, deliver valuable guidance for modulating exciton transport, diffusion and quasiparticle coupling in layered quantum materials, and establish a powerful experimental route to explore low-dimensional exciton physics.

## Main Text

Excitons are bound electron-hole quasiparticles that dominate the optical response of semiconductors and insulators (*1-3*), governing light absorption, emission, and a broad range of many-body optoelectronic phenomena (*4, 5*). Their composite bosonic nature enables collective quantum behavior, including condensation of excitons (*6*) and exciton polaritons (*7*), as well as excitonic superfluidity (*8, 9*). Moiré engineering in two-dimensional (2D) layered systems further enriches correlated excitonic states and other emergent phenomena (*10, 11*), rendering low-dimensional exciton physics a rapidly growing frontier.

A fundamental descriptor underlying these diverse phenomena is the exciton band structure $E(q)$, the energy-momentum relation of the exciton center-of-mass motion. Analogous to electron and phonon dispersions, exciton dispersion $E(q)$ determines the group velocity, effective mass, scattering channels and transport properties, and underpins exciton diffusion (*12, 13*), wave-packet dynamics (*14, 15*), and radiative relaxation (*16-18*). Theoretically, $E(q)$ is highly sensitive to dimensionality: reduced layer thickness modulates dielectric screening, electron-hole exchange, and wavefunction confinement, driving qualitative changes across the bulk-to-2D crossover (*19-21*). In the 2D limit, the long-range electron-hole exchange interaction introduces a nonanalytic contribution to $E(q)$, yielding lightlike, massless dispersion in the long-wavelength limit despite massive charge carriers. This long-wavelength regime is particularly important because it governs radiative coupling, early-time exciton propagation and coherence. Experimentally, however, quantitative, layer-resolved tracking of how dimensionality reshapes $E(q)$ in the long-wavelength limit remains elusive.

Progress in electron energy-loss spectroscopy (EELS) integrated within transmission and scanning transmission electron microscopes (TEM/STEM) has achieved sub-10 meV energy resolution (*22*). Featuring nanoscale spatial resolution and freedom from substrate artifacts, TEM/STEM-EELS offers distinct advantages over resonant inelastic x-ray scattering (RIXS) (*23, 24*) and reflective high-resolution EELS (HREELS) (*25*), making it an exceptionally powerful tool for dispersion measurements. To date, it has been applied to dispersion measurements of a wide range of elementary excitations, including phonons (*26-29*), excitons (*30, 31*), plasmons (*32, 33*), and magnons (*34*). However, typical parallel-beam TEM-EELS only achieves a momentum resolution of ~0.02 $Å^{-1}$ (*30, 35*), while small-convergence-angle STEM-EELS is limited to ~0.1-0.4 $Å^{-1}$ (*28, 31, 34*). Such momentum resolutions are insufficient to resolve the fine structure of exciton dispersion in the long-wavelength limit. It thus remains a major challenge to push the momentum resolution of TEM/STEM-EELS toward the state-of-the-art ~0.002 $Å^{-1}$ attained by RIXS (*36*) and HREELS (*37*).

In this work, we address this challenge by adapting a defocus-engineered STEM-EELS approach (*32, 38*), which has also been referred to as the angle-resolved (AR) mode (*22*), to achieve an ultrahigh momentum resolution of 0.0002 $Å^{-1}$, together with an energy resolution better than 20 meV. This capability enables direct measurements of intrinsic exciton dispersion with unprecedented precision in atomically thin materials. We demonstrate this advance by resolving the dimensionality-dependent exciton dispersion of hexagonal boron nitride (hBN), an ideal van der Waals model system with strong excitonic effects and large oscillator strength (*39*). While theoretical studies have predicted the exciton band structure of hBN, along with the nature of bright and dark excitonic states and their evolution with dimensionality (*20, 40, 41*), experimental efforts have so far only identified direct (*42*) and indirect (*43*) excitons, with merely subtle signatures of linear dispersion emerging in the monolayer limit (*31*). Yet the previous measurement on monolayer hBN (*31*) was limited to a momentum resolution of only ~ 0.24 $Å^{-1}$— far from sufficient to access the small-$q$ linear dispersion regime—much less to quantify the dimensional evolution of the exciton band structure within the ultralow-$q$ region ($q < 0.02$ Å$^{-1}$).

Here, we systematically investigate freestanding hBN samples with varying layer thicknesses, unambiguously mapping the exciton band structure evolution across the 2D-to-3D crossover. With decreasing thickness, the momentum window of the nonanalytic linear-dispersion regime progressively broadens, while its group velocity declines. We further demonstrate that the massless exciton branch can be environmentally tuned by temperature, graphene proximity and interlayer twist. Our results establish a robust microscopy-based route for quantitative quasiparticle band-structure metrology in low-dimensional quantum materials.

**Instrumental access to the ultralow-$q$ regime**

In conventional STEM-EELS experiments (Fig. 1A), a standard convergent-beam configuration employs an electron beam focused onto the sample, where the convergence semi-angle $\alpha$ simultaneously determines the real-space probe size and the spread of incident momenta (*26*). A slot aperture is introduced to enable simultaneous acquisition of momentum-resolved spectral information over a finite $q$ window (*27*). In this geometry, improving momentum resolution requires progressively reducing $\alpha$. In practice, however, the minimum accessible $\alpha$ is constrained by the condenser-lens optics, and further reduction ultimately drives the system toward the parallel-beam TEM-EELS limit (*30*), failing to reach the ultrahigh momentum resolution sought here.

By contrast, the defocus-engineered geometry, a strategy originally introduced by Midgley for high-angular-resolution EELS measurements (*38*), provides an alternative route to an effectively ultralong camera length. In our implementation (Fig. 1B), starting from the original convergent-beam configuration, the projector system is first retuned so that the diverging beam is refocused onto the aperture plane. A controlled defocus $\Delta f$ is then introduced by shifting the sample height, which effectively expands the diffraction pattern at the spectrometer entrance while leaving the size of the probe point essentially unchanged. As a result, the angular resolution is no longer determined solely by $\alpha$, but is approximately given by $d/\Delta f$ (where $d$ is the probe size set by $\alpha$). Therefore, a larger convergence semi-angle $\alpha$, which gives a smaller probe size $d$, together with a larger defocus $\Delta f$, can in principle yield a better momentum resolution. Further details of the optical geometry, momentum calibration and resolution estimate are provided in Materials and Methods and figs. S1 and S2.

Figure 1C summarizes the calculated and experimental momentum resolutions for the two different strategies. The gray line shows the dependence of the momentum resolution on the convergence semi-angle $\alpha$ in the conventional convergent-beam configuration. The minimum practically accessible convergence semi-angle in this work, $\alpha = 0.3$ mrad, is marked as point A and corresponds to a momentum resolution of 0.027 Å$^{-1}$. By contrast, the defocus-engineered geometry provides a continuously tunable route to substantially improved momentum resolution. At a fixed convergence semi-angle of 8 mrad, the effective momentum resolution is governed by the defocus $\Delta f$, as shown by the orange curve in Fig. 1C. The experimental resolutions (orange spots) obtained from repeated calibrations agree well with the calculated trend, demonstrating that the accessible momentum resolution can be tuned systematically down to $2 \times 10^{-4}$ Å$^{-1}$. A more direct visualization of the improvement is provided in Fig. 1D, where the effective angular point-spread functions under the three representative settings [A ($\alpha = 0.3\,\text{mrad},\ \Delta f = 0$), B ($\alpha = 8\,\text{mrad},\ \Delta f = 20\,\mu\text{m}$), and C ($\alpha = 8\,\text{mrad},\ \Delta f = 100\,\mu\text{m}$) in Fig. 1C] are plotted. Relative to the broad angular distribution in the conventional $\alpha = 0.3$ mrad condition, the defocus-engineered geometry compresses the effective angular spread dramatically, reaching $\sim 43$ µrad at $\Delta f = 20$ µm and $\sim 8.5$ µrad at $\Delta f = 100$ µm, thereby entering the microradian regime. The real-space scan images shown alongside the profiles illustrate the corresponding spatial resolution under these representative settings. Although the trade-off between spatial and momentum resolution still remains, the controlled defocus provides a flexible experimental knob that allows rapid switching between different operating modes according to the measurement target. Notably, with a larger convergence semi-angle $\alpha$ (corresponding to a smaller probe size $d$) and a larger defocus value $\Delta f$, even higher momentum resolution should in principle be achievable, although its practical implementation may be constrained by aberrations and alignment stability and specimen geometry.

This improvement in momentum resolution has a direct impact on what can be resolved experimentally. As a representative example, we examine the excitonic response of 15-layer hBN under different experimental configurations to illustrate the transition from qualitative observation to quantitative access in the long-wavelength limit. Under the conventional $\alpha = 0.3$ mrad setting, the momentum-tagged energy distribution curves (EDCs) show separated exciton and $\pi$ plasmon features, as indicated by the red and purple dashed lines in Fig. 1E. However, the exciton peak remains strongly broadened in the vicinity of Γ point, and its behavior in the ultralow-$q$ regime cannot be extracted with sufficient reliability for quantitative analysis. This limitation becomes evident in the dispersion series shown in Fig. 1F. The first slice (from the left), corresponding to the dispersion map associated with Fig. 1E, shows that the information near the zone center is largely obscured by finite momentum broadening. Once the defocus-engineered geometry is introduced with improved momentum resolution, the exciton branch evolves from a broadened spectral feature into a well-defined dispersive mode over a much narrower momentum interval with increasing $\Delta f$. In the highest-resolution condition shown here, $\Delta f = 400$ µm, the momentum-tagged EDCs in Fig. 1G allow the exciton peak positions to be tracked robustly over the ultralow-momentum range below 0.016 $\text{Å}^{-1}$. This capability provides the experimental foundation for the layer-by-layer determination of the nonanalytic ultralow-$q$ exciton dispersion presented below.

**Layer-by-layer evolution of exciton band structure in freestanding hBN**

We turn our attention to monolayer hBN, in which the longitudinal singlet exciton is expected to exhibit a lightlike, or effectively massless, dispersion in the long-wavelength limit (*31*). Under the conventional $\alpha = 0.3$ mrad condition, the dispersion map acquired along the M-Γ-M direction shown in Fig. 2A, together with the corresponding EDCs shown in Fig. 2B, already reveals a well-defined exciton branch spanning approximately 6.3 to 8.2 eV, together with a higher-energy plasmon branch extending from about 7 to 13 eV. However, although the momentum resolution achieved in our conventional convergent-beam configuration already surpasses that of previous experiments (*31*), a clear and quantitative analysis remains challenging. In particular, the dispersion within the expected nonanalytic regime still appears predominantly parabolic, indicating that the experimentally accessible momentum resolution remains insufficient to fully resolve the intrinsic low-$q$ behavior predicted by theory. Once the defocus-engineered geometry is applied, even at a moderate defocus of $\Delta f = 20$ µm, the characteristic V-shaped exciton dispersion at low $q$ in monolayer hBN becomes directly resolved in the dispersion map shown in Fig. 2C, and the corresponding momentum-tagged EDCs in Fig. 2D reveal a clear and systematic linear shift of the exciton peak with momentum. These results demonstrate that the defocus-engineered strategy, through its substantial enhancement of momentum resolution, enables the direct and unambiguous observation of the massless linear exciton dispersion in monolayer hBN.

We then extend the measurement to a thickness series of freestanding hBN with L = 1, 2, 4, 5, 15, and 25 layers (see fig. S3) under the same acquisition condition ($\Delta f = 100$ μm) and over the same momentum window of $\pm 0.064$ Å$^{-1}$. As shown in Fig. 2E, a systematic layer-dependent evolution of the exciton band structure is directly observed. In the monolayer, the exciton branch retains its linear dispersion across the entire displayed momentum range. With increasing thickness, the momentum range over which this nonanalytic, nearly linear behavior persists progressively shrinks toward $q = 0$, while the branch becomes correspondingly steeper. By 15 and 25 layers, this ultralow-$q$ linear feature is confined to a much narrower momentum interval, although it remains directly resolvable. The data-processing procedure used for subsequent quantitative analysis is shown in figs. S4 to S6. For reference, a bulk-limit result is presented in fig. S7, where the exciton dispersion is well described by a parabolic form. The corresponding EDCs in Fig. 2F provide the spectral details of the evolution of the exciton peak trajectory from the extended linear regime in the monolayer to a narrower nonanalytic feature in thicker samples.

The well-resolved monolayer exciton branch in Fig. 2C and the first panel of Fig. 2E enables a reliable linear fit along Γ-M, yielding a group velocity of $v_g = 2.0 \times 10^{-3}\ c$, where $c$ is the speed of light in vacuum. This velocity is lower than the previously reported value (*31*). As illustrated in Fig. 2A and fig. S5A, insufficient momentum resolution can cause the exciton signal to overlap with phonon-sideband (*35*) or high-energy plasmonic spectral weight, biasing the extracted peak position toward higher energy and leading to an overestimated apparent slope. By directly resolving the ultralow-$q$ exciton branch, the defocus-engineered geometry removes this ambiguity and enables quantitative analysis. Additional comparisons are shown in fig. S8.

**Dimensional dielectric screening and environmental control**

To interpret the thickness-dependent trends in Fig. 2, E and F, we first consider how dielectric screening evolves with thickness. As illustrated schematically in Fig. 3A, reducing thickness weakens dielectric screening because the electric-field lines associated with an electron-hole excitation are less confined within the crystal and spill more strongly into the surrounding vacuum. As a result, atomically thin insulators and semiconductors exhibit dielectric responses that evolve systematically, and increasingly nonlocally, from the few layers to monolayer (*44, 45*). For neutral electron-hole excitations, this dimensional reduction of dielectric screening is most consequential in the long-wavelength limit, where both the screened Coulomb interaction and the exchange component of the electron-hole interaction are especially sensitive to thickness (*46, 47*). In this regime, reduced dimensionality induces strong nonanalytic renormalization of the exciton energy near $q = 0$, thereby reshaping the exciton band structure (*19-21*).

The extracted dispersion points are summarized in Fig. 3B to quantitatively compare the exciton dispersions across different thicknesses. With increasing thickness, the low-$q$ dispersion becomes progressively steeper, while the onset of curvature shifts systematically to smaller momenta. Although Fig. 3B focuses on the low-$q$ window used for fitting, the monolayer already exhibits an extended nearly linear branch that persists up to 0.2 Å$^{-1}$, as seen in Fig. 2C. This evolution indicates that enhanced dielectric screening does not remove the nonanalytic feature of the exciton band, but instead confines the linear region to an increasingly narrow momentum range. To quantify this evolution, we fitted the extracted dispersions using a finite-thickness quasi-2D exchange model (*20*)

$$E(q) = E_0 + A\left[1 - \frac{q_c}{q}\left(1 - e^{-q/q_c}\right)\right], \tag{1}$$

where $E_0$ is the exciton energy at $q = 0$, corresponding to the band gap reduced by the exciton binding energy. The amplitude $A$ is proportional to the squared projected optical transition dipole matrix elements. The characteristic momentum $q_c$ defines the crossover momentum from the 2D long-wavelength nonanalytic regime to the finite-thickness regime, and is expected to scale inversely with the effective thickness of the electronic charge distribution. As shown by the solid lines in Fig. 3B, this model captures the experimental dispersion trends well across the thickness series.

The extracted fitting parameters are summarized in Fig. 3C. Consistent with the evolution of the dispersions discussed above, $q_c$ decreases rapidly with increasing layer number, from $1.82 \times 10^{-1}$ Å$^{-1}$ in the monolayer to $3.0 \times 10^{-3}$ Å$^{-1}$ in 25L, reflecting the progressive compression of the nonanalytic regime as dielectric screening becomes stronger. We further find that the product $q_c t$, where $t$ is the sample thickness, remains of order unity and tends towards saturation at larger thicknesses. This near-unity scaling behavior highlights the direct role of dimensionality in governing the exciton dispersion. The low-$q$ group velocity, $v_g = (1/\hbar)(dE/dq)_{q\to 0}$, increases systematically with thickness, from $2.0 \times 10^{-3} c$ in monolayer hBN to $2.9 \times 10^{-2}\ c$ for 25L. In the monolayer limit, this value together with the linear branch extending up to 0.2 Å$^{-1}$ provides direct evidence for the massless exciton expected in the strict 2D limit. We also examined an alternative model in which the out-of-plane component of the electronic wavefunction decays exponentially rather than being uniformly confined within a finite slab (*20*). This analysis yields a decay length $\lambda$ that increases from ~2.5 Å in 1L to ~187 Å in 25L, with $\lambda/t$ remaining of order unity (~0.8-2.7). Together, these results provide the first experimental observation and quantitative analysis of the dimensional evolution of exciton dispersion. Further details are provided in fig. S9.

Having established the dimensional evolution of the exciton band structure, we further examined how to modulate the massless exciton in monolayer hBN by changing the external environments. Figure 4, A to H, shows the dispersion maps of freestanding monolayer hBN at 300 K and 100 K (Fig. 4, A and B), coupled to one and two graphene layers (Fig. 4, C and D), and a twist-angle series with $\theta$ = 3.8°, 11.4°, 23.1°, 30.0° (Fig. 4, E to H). In all cases, the long-wavelength exciton retains a linear dispersion, indicating that the massless behavior is robust against temperature variation, dielectric proximity and interlayer registry. However, the group velocity $v_g$ and the optical gap $E_0$ show environment-dependent variations, as quantified by the linear fits in Fig. 4I and summarized in Fig. 4J. Graphene-coupled samples show additional low-energy spectral weight from graphene-related $\pi$ plasmon and interband transition losses, confirming the presence of the adjacent graphene layers. Nevertheless, the hBN exciton branch remains well resolved and preserves its linear low-$q$ dispersion with modest changes in $v_g$ and $E_0$, consistent with the near-unity real part of the dielectric function of graphene in the deep-ultraviolet range (*48*).

The temperature-dependent measurements reveal a pronounced reduction of $E_0$ at 100 K relative to the room-temperature results. The energy difference is comparable to the optical-phonon energy of hBN, for which phonon replicas are typically separated from the excitonic transition by approximately 0.17 eV. Therefore, the lower $E_0$ observed at 100 K suggests a reduced contribution from phonon-assisted spectral weight and a closer approach to the underlying zero-phonon excitonic branch, whereas the room-temperature spectrum likely contains stronger convolution with phonon replicas and thermally broadened exciton-phonon self-energy (*49, 50*). The concurrent change in the fitted $v_g$ is consistent with a $q$-dependent exciton-phonon self-energy that modifies the apparent direct-exciton dispersion. These results provide a momentum-resolved signature of exciton-phonon coupling in monolayer hBN.

**Discussion**

In summary, using an ultrahigh-momentum-resolution configuration, we conducted a systematic study on the dimensional effect of hBN exciton dispersion at the low-$q$ region. First, we provide unprecedentedly high-precision experimental evidence that excitons in monolayer hBN exhibit a massless linear dispersion in the ultralow-$q$ regime, in sharp contrast to the conventional parabolic behavior observed in the bulk limit. Second, by systematically tracking the dimensional evolution of the exciton band structure, we reveal that the linear excitonic branch persists beyond the strict 2D limit and survives throughout a finite low-dimensional regime. The characteristic momentum scale associated with the nonanalytic linear dispersion is further found to exhibit an inverse scaling relation with sample thickness. Furthermore, we show that the group velocity of the massless excitonic mode can be engineered through dimensional and environmental tuning, highlighting new opportunities for manipulating collective excitonic transport and optoelectronic functionalities in van der Waals materials.

More broadly, our results demonstrate the capability of the defocus-engineered STEM-EELS strategy to attain ultrahigh momentum resolution and directly probe emergent long-wavelength collective excitations in quantum materials. Long-range Coulomb interactions become increasingly dominant in the long-wavelength limit, making many-body screening (*51*), collective charge fluctuations (*52, 53*), and quantum critical dynamics (*54, 55*) far more pronounced at ultralow-$q$. Access to this momentum regime is therefore crucial for uncovering the fundamental dynamic response of quantum matter. The substantial enhancement in momentum resolution achieved here opens a direct experimental pathway toward exploring emergent collective phenomena.

## Materials and Methods

1. The preparation of freestanding hBN samples

Monolayer single-crystal hBN was grown on CuNi(111)/sapphire wafers by chemical vapour deposition (CVD). A 500-nm-thick CuNi film with a Ni concentration of 20% was first deposited on single-crystal sapphire by co-sputtering physical vapour deposition. To obtain a single-crystal CuNi(111) substrate, the CuNi film was annealed at 1000 °C for 2 hours under ambient pressure in a tube furnace (Thermo Scientific) with a gas flow of 500 sccm Ar and 10 sccm $H_2$. Ammonia borane (97%, Sigma Aldrich) was used as the precursor and placed in a quartz tube upstream of the CVD system. The CuNi(111) substrate was heated to 1000 °C under ambient pressure in 500 sccm Ar and 100 sccm $H_2$. The pressure was then reduced to 100 Torr by introducing 500 sccm Ar and $H_2$. The precursor was heated to 100 °C using a heating tape, and hBN growth proceeded for approximately 30 min. After growth, all gas flows were immediately turned off, and the sample was allowed to cool to room temperature. To transfer hBN from the CuNi(111)/sapphire wafer onto TEM grids, a batch of TEM grids was placed directly onto the wafer. Several drops of isopropanol were then added to cover the surface of the TEM grids. After the solvent had volatilized, the TEM grids adhered to the hBN layer. The resulting stack of TEM grids/hBN/CuNi(111)/sapphire was fully immersed in a 1 M $(NH_4)_2S_2O_8$ water solution to digest the underlying 500-nm CuNi(111) film. This etching step yielded TEM grids covered by a continuous hBN film suspended over the porous carbon film.

Multilayer hBN flakes were mechanically exfoliated from bulk single crystals onto Si/$SiO_2$ substrates. Flakes with the desired layer numbers were identified by atomic force microscopy (AFM). A drop of isopropanol was placed on the flakes, and a TEM grid was gently laid onto the droplet. As the isopropanol evaporated, the grid came into contact with the flakes. Dilute HF solution (1.5%) was then introduced at the edge of the grid to etch the $SiO_2$ layer. The detached grid carrying multilayer hBN flakes was collected, thoroughly rinsed with deionized water and dried.

The hBN/graphene heterostructures were prepared by transferring monolayer hBN onto homemade monolayer graphene TEM grids using a PMMA-assisted wet-transfer process. Monolayer graphene was first grown on commercial Cu foil by CVD. The Cu foil was heated to 1030 °C under 500 sccm Ar and 10 sccm $H_2$, followed by the introduction of 0.1 sccm $CH_4$ during growth. After growth, the CVD system was cooled to room temperature. To prepare graphene TEM grids, PMMA was spin-coated onto the graphene/Cu foil and baked at 120 °C for 3 min. The Cu foil was then etched away in 4% $(NH_4)_2S_2O_8$ solution. The floating PMMA/graphene film was picked up with a commercial holey-carbon-film TEM grid and baked at 120 °C for 10 min to improve adhesion between graphene and the carbon support. The PMMA layer was subsequently removed in acetone. Monolayer hBN was then transferred onto the graphene-covered TEM grids using the same PMMA-assisted transfer procedure, yielding freestanding hBN/graphene heterostructures supported by the holey carbon film.

2. EELS and imaging experiments

STEM-EELS experiments were performed on a Nion U-HERMES100S instrument equipped with a monochromator and aberration correctors, operated at an accelerating voltage of 30 kV. Low-temperature measurements were carried out using a liquid-nitrogen cooling holder (NNM6-ST, Zeptools Technology Co., Ltd). A hybrid-pixel direct-detection camera (Dectris ELA) was used for data collection. Defocus-engineered EELS datasets were acquired with a convergence semi-angle of $\alpha$ = 8 mrad and a spectral sampling of 10 meV per channel over 1028 channels. For conventional small convergence angle measurements acquired with $\alpha$ = 0.3 mrad, the spectral sampling was 16 meV per channel. The monochromated probe current at the sample was approximately 0.3 pA, corresponding to an initial beam current of approximately 3.8 pA before monochromation. Momentum-resolved spectra were collected using a 2 mm × 0.125 mm slot aperture, yielding 423 effective momentum pixels collected on the detector. The typical full width at half maximum of the zero-loss peak was approximately 20 meV. Each dataset was acquired with a 10 s exposure and averaged over 400 spectra. HAADF images were acquired with a convergence semi-angle of $\alpha$ = 23 mrad and a detector collection-angle range of $\beta$ = 80-210 mrad.

3. Momentum resolution estimation and momentum calibration

In the defocus-engineered geometry, the ideal angular resolution can be simply estimated as $\delta\theta_{\mathrm{ideal}} = d/|\Delta f|$, where $d$ is the diffraction-limited probe size and $\Delta f$ is the sample-height defocus, as illustrated in fig. S1. For the 30 kV, $\alpha = 8$ mrad condition used in this work, $d$ was estimated as $0.61\lambda_e/\sin\alpha$, where $\lambda_{\mathrm{e}}$ (= 6.98pm) is the relativistic electron wavelength, giving $d = 0.53$ nm. The corresponding ideal momentum resolution is therefore $\delta q_{\mathrm{ideal}} = k_0 \delta\theta_{\mathrm{ideal}} = (2\pi/\lambda_e) d/|\Delta f|$, where $k_0 = 2\pi/\lambda_{\mathrm{e}}$ is the electron wave vector. This estimate was used to generate the ideal momentum-resolution curve shown in Fig. 1C.

In practice, the actual momentum resolution is broader than this ideal estimate because of aberrations, source-size broadening and other instrumental effects. We therefore further determined the experimental momentum resolution for each configuration using the calibration procedure described below, which gives the data points shown in Fig. 1C.

After setting the projector lens and the sample-height defocus, the electron beam was first parked at the center of the scan field. We recorded the beam position at the Ronchi CCD camera and at the EELS aperture plane. The beam width on the Ronchi CCD, denoted as $\gamma$, was used to estimate the angular broadening of the incident beam. The length of the slot aperture was measured at the EELS aperture plane and denoted as $L_{\mathrm{s}}$, in arbitrary detector units. To calibrate the angular scale, the descan coils were switched off and the incident beam was displaced by a known distance $\Delta x$ away from the scan center. Under a sample-height defocus $\Delta f$, this real-space displacement corresponds to an angular change of approximately

$$\Delta\theta = \frac{\Delta x}{|\Delta f|}.$$

The corresponding beam displacement was measured as $\Delta L_{\mathrm{r}}$ on the Ronchi CCD and as $\Delta L_{\mathrm{e}}$ at the EELS aperture plane. The angular scale on the Ronchi CCD is therefore

$$\frac{\Delta\theta}{\Delta L_{\mathrm{r}}} = \frac{\Delta x}{|\Delta f| \Delta L_{\mathrm{r}}}.$$

The angular resolution was estimated from the full width at half maximum spot size $\gamma$ on the Ronchi CCD as

$$\delta\theta = \frac{\gamma}{\Delta L_{\mathrm{r}}} \frac{\Delta x}{|\Delta f|}.$$

Accordingly, the momentum resolution is

$$\delta q = k_0 \delta\theta = \frac{2\pi}{\lambda_{\mathrm{e}}} \frac{\gamma}{\Delta L_{\mathrm{r}}} \frac{\Delta x}{|\Delta f|},$$

The momentum scale of the EELS datasets was calibrated in the same way using the beam displacement measured at the EELS aperture plane. The total momentum window collected through the slot aperture is

$$\Delta q_{\mathrm{span}} = k_0 \frac{L_{\mathrm{s}}}{\Delta L_{\mathrm{e}}} \frac{\Delta x}{|\Delta f|} = \frac{2\pi}{\lambda_{\mathrm{e}}} \frac{L_{\mathrm{s}}}{\Delta L_{\mathrm{e}}} \frac{\Delta x}{|\Delta f|}.$$

Here, $L_{\mathrm{s}}$ and $\Delta L_{\mathrm{e}}$ can be measured in arbitrary detector units, because only their ratios enter the calibration.

In our experiment, for $\Delta f = 400\ \mu\mathrm{m}$, a beam displacement of $\Delta x = 100$ nm produced $\Delta L_{\mathrm{r}} = 0.09263$ a.u. on the Ronchi CCD, while the measured beam broadening was $\gamma = 0.00079$ a.u. This gives an angular resolution of $2.1\ \mu\mathrm{rad}$, corresponding to a momentum resolution of $1.9 \times 10^{-4}$ Å$^{-1}$ at 30 kV. The slot length at the EELS aperture plane was $L_{\mathrm{s}} = 136.5$ a.u., and the beam displacement at the EELS aperture plane was $\Delta L_{\mathrm{e}} = 72.3$ a.u. This gives a total collected momentum window of $0.043$ Å$^{-1}$. For a 423-pixel momentum axis, the calibrated momentum sampling is therefore $1.00 \times 10^{-4}$ Å$^{-1}$ per pixel.

4. EELS data processing

Custom-written MATLAB codes were used to process all acquired EELS datasets. For each *q*-EELS dataset, the 400 repeated acquisitions were first aligned using normalized cross-correlation to correct for possible beam-energy drift and momentum-axis distortions, and were then summed to improve the signal-to-noise ratio. To better visualize the dispersive features, each momentum-resolved spectrum was further normalized by its integrated intensity within the displayed energy window, so that the total counts were comparable across different momenta. The normalized EELS maps were Gaussian-smoothed for display. The peak positions of the exciton and plasmon modes were extracted by fitting the corresponding energy-loss spectra with Gaussian functions.

5. GW-BSE calculations of monolayer hBN

Finite-momentum exciton dispersions of monolayer hBN in fig. S8 were calculated following the GW-BSE workflow of *Liu et al.* (*31*). Ground-state Kohn-Sham wavefunctions were obtained with density functional theory using the Quantum ESPRESSO package (*56*), ONCV norm-conserving pseudopotentials (*57*), and the PBE exchange-correlation functional (*58*). The experimental in-plane lattice constant was used, with a 20 Å vacuum spacing along the out-of-plane direction. The plane-wave kinetic-energy cutoff was 70 Ry, and the ground-state Brillouin zone was sampled on a 24 × 24 × 1 k grid.

Quasiparticle corrections were computed with BerkeleyGW (*59*) within the Hybertsen-Louie generalized plasmon-pole GW approach (*60*). The dielectric matrix was evaluated with a 30 Ry cutoff and 2000 empty bands together with slab Coulomb truncation to remove spurious image interactions. Finite-Q exciton energies were obtained by solving the Bethe-Salpeter equation in the Tamm-Dancoff approximation for exciton center-of-mass momenta Q along Γ-M, with the long-range exchange interaction retained. The finite-Q BSE kernel was evaluated on a uniform 24× 24 × 1 k grid without interpolation, using a reduced dielectric cutoff of 10 Ry. The active transition space included the two highest valence bands and two lowest conduction bands. The calculated low-energy branches were mirrored to represent the M-Γ-M path. For comparison with experiment, all calculated exciton energies were shifted by a single constant to align the calculated Γ-point exciton energy with the measured value; no additional fitting was applied to the calculated dispersion.

**Acknowledgments:**

**Funding:** The work was supported by the National Natural Science Foundation of China (Grant No. 52125307 to P.G., No. 12504198 to J.L. and No. 52188101 to X.C.). P.G. acknowledges the support from the Xplorer Prize.

**Author contributions:** P.G. conceived the project. P.H. and Jiade Li performed the STEM-EELS experiments, with guidance from N.D. and T.L. on the electron-optical configuration. P.H. performed the data processing and analysis with assistance from Jiade Li. Jiangxu Li provided theoretical and computational support with assistance from X.C. Jiakai Wang prepared the multilayer samples under the guidance of Y.S. and Y.Y. Jiahao Wang and W.S. prepared the monolayer samples under the guidance of H.P. Q.G. prepared the BN/graphene twisted heterostructures under the guidance of K.L. B.H. and R.S. contributed to discussions of the electron-optical setup and data interpretation. P.H. and Jiade Li wrote the manuscript under the supervision of P.G. All authors discussed the results and contributed to the manuscript.

**Data, code, and materials availability:** All data are available in the main text or the supplementary materials. The code for this paper is available from the corresponding authors upon reasonable request. Details for the synthesis and characterization of all materials are in the materials and methods.

**Competing interests:** The authors declare no competing interests.

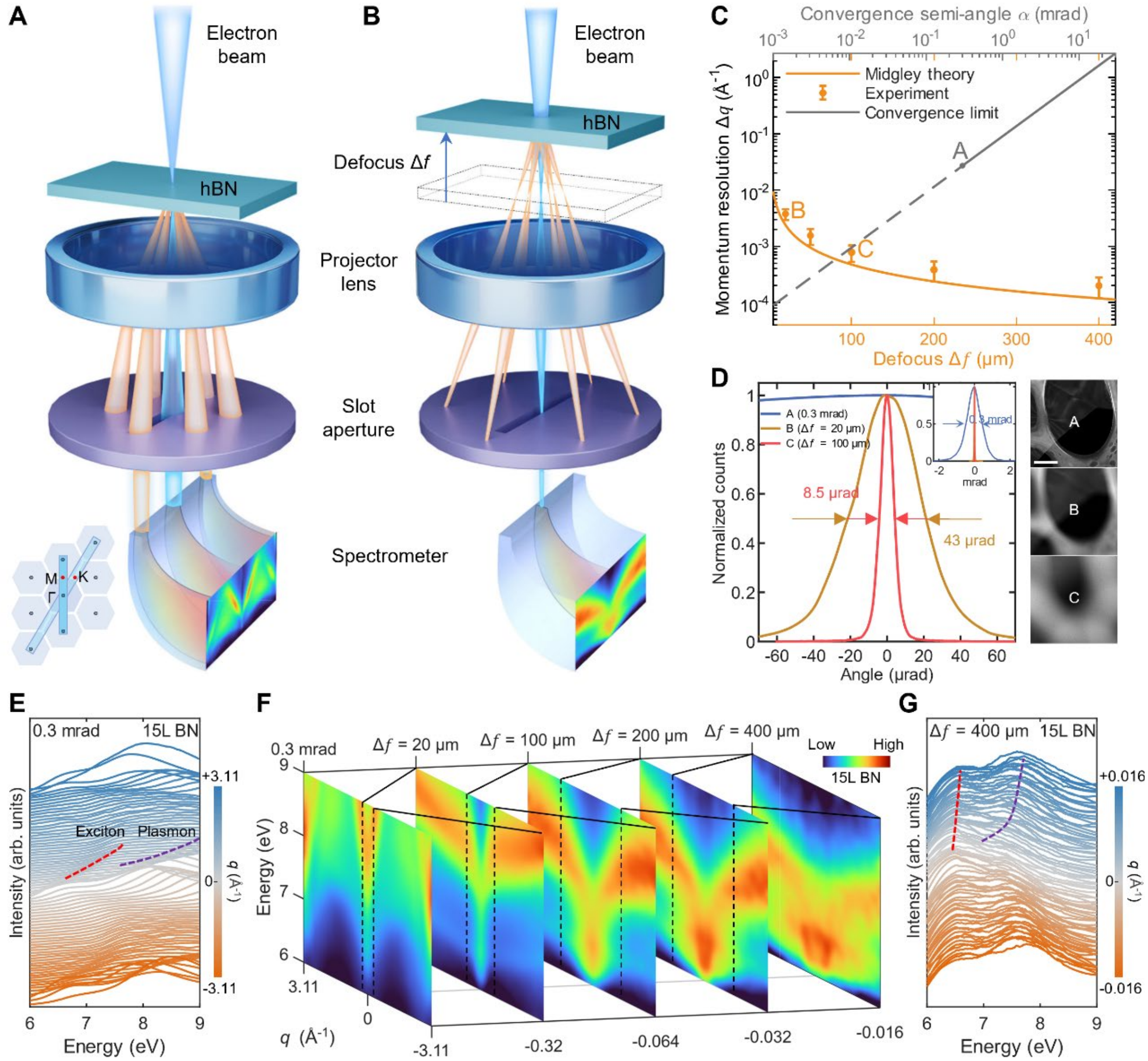

**Fig. 1. Experimental setup for high-momentum-resolved STEM-EELS.** (**A**) Schematic of conventional momentum-resolved STEM-EELS using a small convergence semi-angle. The lower-left inset shows the Brillouin zone of hBN and the slot aperture used to select specific momentum directions. The map after the spectrometer shows a full-Brillouin-zone dataset acquired along the Γ-M direction with a convergence semi-angle of 0.3 mrad, spanning an energy range from 5.7 to 13 eV. (**B**) Defocus-engineered geometry: retuning the projector lens to refocus the diffraction pattern onto the slot aperture, and a controlled defocus $\Delta f$ maps scattering angle to a spatial displacement at the spectrometer entrance, yielding greatly enhanced momentum resolution. The map after the spectrometer shows a dataset acquired along the Γ-M direction with an 8 mrad convergence semi-angle and $\Delta f$ = 200 μm, covering a momentum range of $\pm$0.032 Å$^{-1}$ and an energy range from 6 to 9 eV. (**C**) Calculated and experimental momentum resolutions. The gray curve denotes the conventional convergence-limited resolution, whereas the orange curve denotes the ideal defocus-engineered resolution calculated for an 8 mrad convergence semi-angle. Symbols denote experimentally measured momentum resolutions. Labels A-C indicate three representative settings: A, $\alpha$ = 0.3 mrad; B, $\Delta f$ = 20 μm; and C, $\Delta f$ = 100 μm. (**D**) Effective angular point-spread functions for representative conditions. Inset: the broader angular distribution

at $\alpha$ = 0.3 mrad for comparison. The right panel shows representative real-space scan high-angle annular dark-field (HAADF) images acquired under the corresponding conditions. Scale bar, 1 µm. (**E**) Momentum-tagged energy-distribution curves (EDCs) acquired for 15-layer hBN under the $\alpha$ = 0.3 mrad condition. Red and purple dashed lines are guides to the eye, indicating the exciton and plasmon features, respectively. (**F**) $\omega$-$q$ slices acquired under different acquisition conditions. (**G**) Momentum-tagged EDCs acquired for 15-layer hBN under the $\Delta f$ = 400 µm condition.

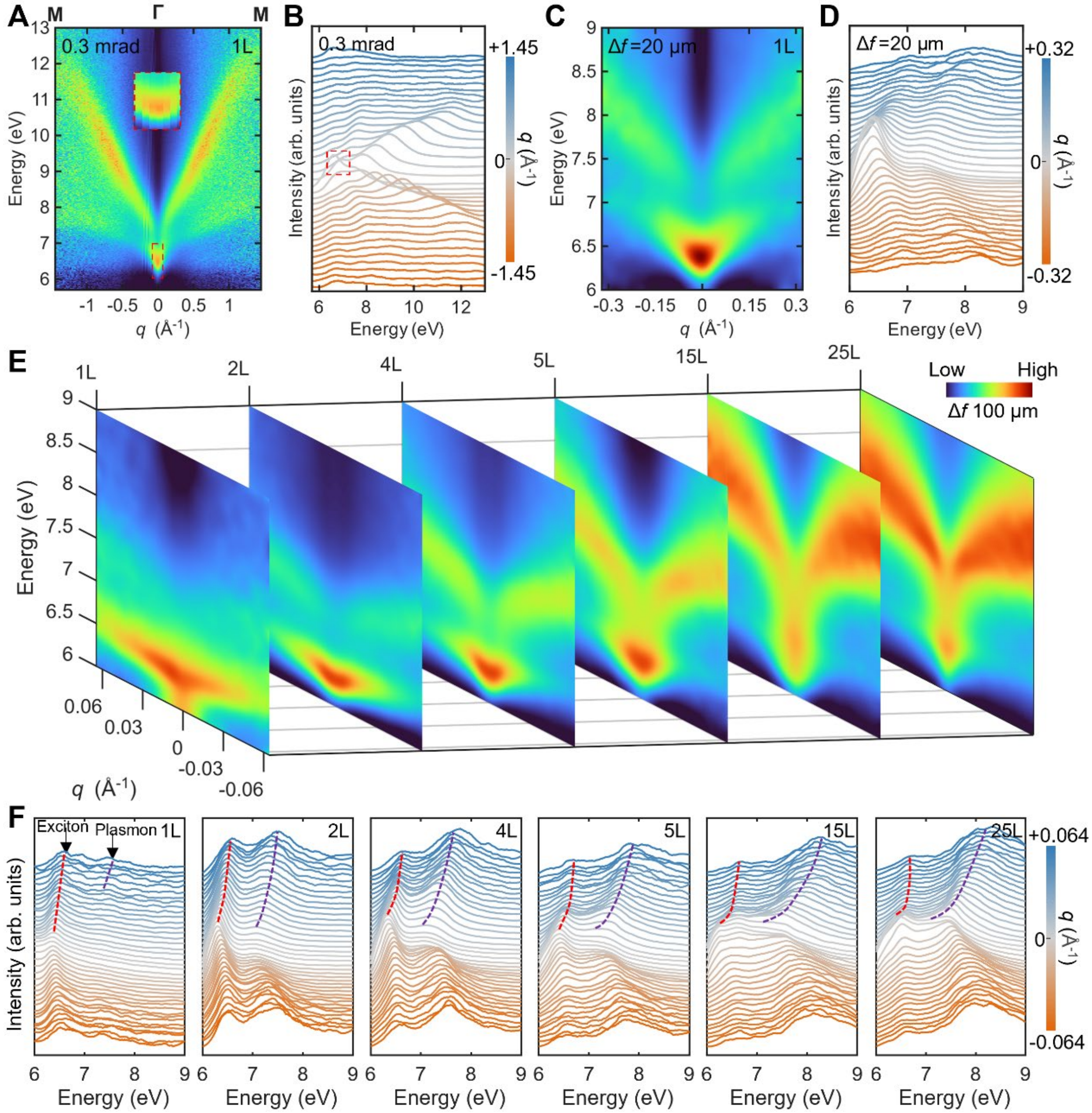

**Fig. 2. Thickness evolution of exciton band structure in freestanding hBN.** (**A** and **B**) Conventional STEM-EELS results on monolayer hBN ($\alpha$ = 0.3 mrad): $\omega$-$q$ map along the M-Γ-M (A) and the corresponding momentum-tagged EDCs (B). (**C** and **D**) Defocus-engineered STEM-EELS results on monolayer hBN ($\Delta f$ = 20 μm): $\omega$-$q$ map (C) and momentum-tagged EDCs (D). (**E**) Thickness series (L = 1, 2, 4, 5, 15, 25 layers) of $\omega$-$q$ maps acquired at $\Delta f$ = 100 μm. (**F**) Corresponding momentum-tagged EDCs for each thickness. Red and purple dashed lines are guides to the eye, indicating the exciton and plasmon features, respectively.

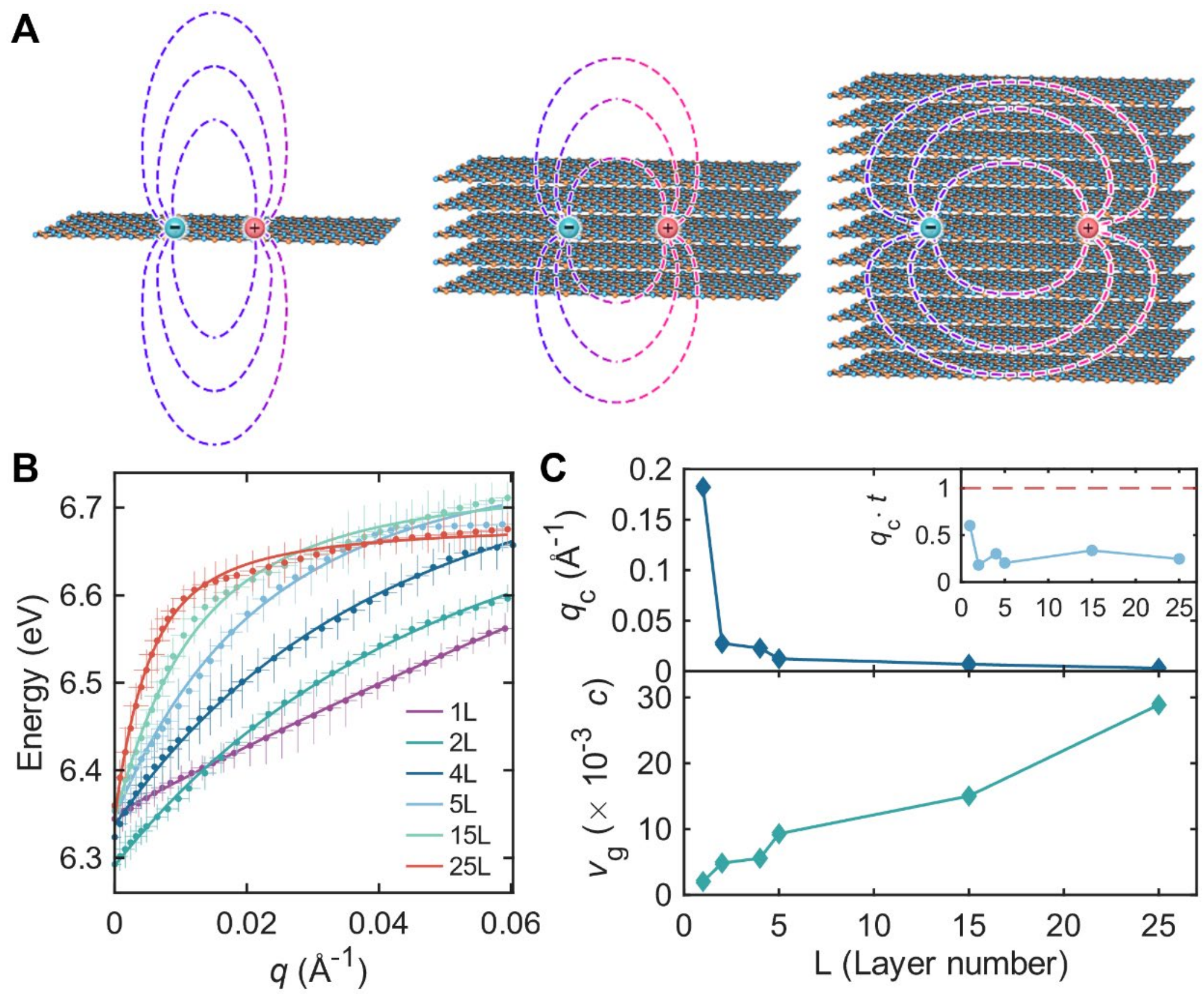

**Fig. 3. Screening-controlled crossover of exciton band structure.** (**A**) Conceptual schematic illustrating the evolution from weak to strong dielectric screening as the thickness increases from monolayer to bulk limit. Dashed contours schematically depict the electric-field lines associated with the electron-hole pair. (**B**) Extracted exciton dispersions, $E(q)$, for representative thicknesses (1L, 2L, 4L, 5L, 15L and 25L). Filled symbols represent the experimental data extracted from the spectra, and solid curves are the corresponding fits using Eq. (1). Horizontal and vertical error bars denote the uncertainties in momentum and energy, respectively, estimated from the instrumental resolution together with peak-finding errors. (**C**) Thickness dependence of the characteristic crossover momentum, $q_c$, and the exciton group velocity $v_g$ extracted from fits in (B). Inset: fitted screening parameter $q_c t$ as a function of layer number L, where $t$ is the sample thickness.

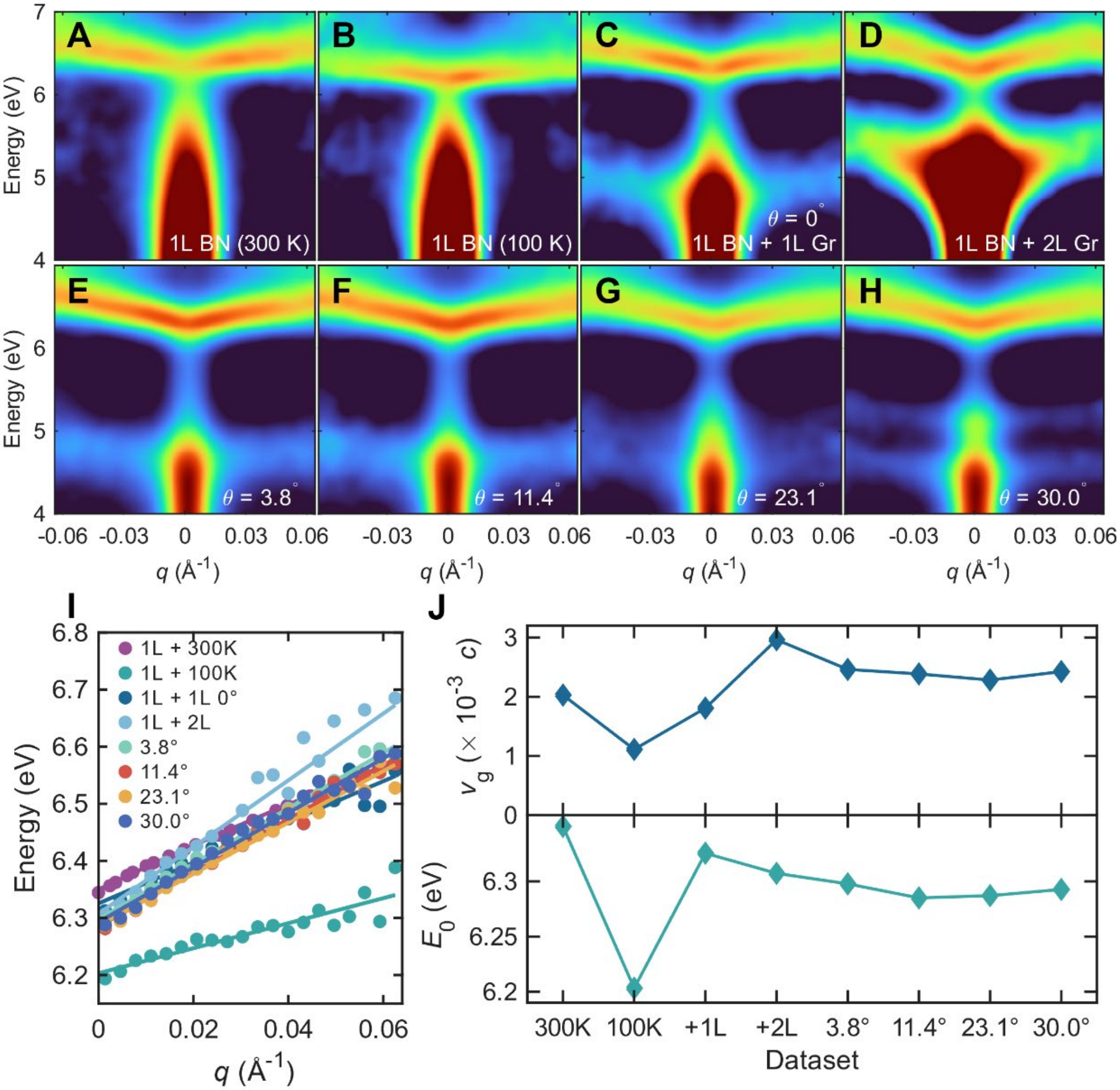

**Fig. 4. Environmental control of massless exciton.** (**A** to **H**), $\omega$-$q$ maps for freestanding monolayer hBN acquired at $\Delta f$ = 100 μm: 300 K (A), 100 K (B), coupled to one graphene layer (C) and to two graphene layers (D), and a twist-angle series ($\theta$ = 3.8°, 11.4°, 23.1°, 30.0°) (E to H). Panels (C) to (H) were measured at 300 K. (**I**) Extracted exciton dispersions, E($q$), from the datasets in (A) to (H) (small dots), together with linear fits (solid lines). (**J**) Summary of the extracted group velocity $v_g$ (top) and optical gap $E_0$ (bottom) for each dataset.